# A PSEUDO EMPIRICAL LIKELIHOOD APPROACH FOR STRATIFIED SAMPLES WITH NONRESPONSE

By Fang Fang, Quan Hong and Jun Shao[1]

*University of Wisconsin-Madison, Eli Lilly and Company and University of Wisconsin-Madison*

Nonresponse is common in surveys. When the response probability of a survey variable $Y$ depends on $Y$ through an observed auxiliary categorical variable $Z$ (i.e., the response probability of $Y$ is conditionally independent of $Y$ given $Z$), a simple method often used in practice is to use $Z$ categories as imputation cells and construct estimators by imputing nonrespondents or reweighting respondents within each imputation cell. This simple method, however, is inefficient when some $Z$ categories have small sizes and ad hoc methods are often applied to collapse small imputation cells. Assuming a parametric model on the conditional probability of $Z$ given $Y$ and a nonparametric model on the distribution of $Y$, we develop a pseudo empirical likelihood method to provide more efficient survey estimators. Our method avoids any ad hoc collapsing small $Z$ categories, since reweighting or imputation is done across $Z$ categories. Asymptotic distributions for estimators of population means based on the pseudo empirical likelihood method are derived. For variance estimation, we consider a bootstrap procedure and its consistency is established. Some simulation results are provided to assess the finite sample performance of the proposed estimators.

**1. Introduction.** Nonresponse is a common phenomenon in sample surveys. Let $Y$ be a variable of interest in a survey. The probability of having a nonrespondent in $Y$ typically depends on the unobserved value of $Y$, which creates a great challenge in the analysis of incomplete survey data. A common approach is to assume that the dependence of the nonresponse probability on $Y$ is through an auxiliary categorical variable $Z$

Received February 2007; revised November 2007.
[1]Supported in part by NSF Grants DMS-04-04535 and SES-07-05033.

whose values are observed for all sampled units in the survey. More precisely, $P(\delta=1|Y,Z) = P(\delta=1|Z)$, where $\delta$ is the response indicator for $Y$. That is, conditional on $Z$, $\delta$ and $Y$ are statistically independent. This response mechanism is referred to as the unconfounded response mechanism by Lee, Rancourt and Särndal (1994), which is the same as missing at random (MAR) [Rubin (1976)]. In the 2002 Survey of Industrial Research and Development (SIRD), for example, a variable $Y$ with nonresponse can be one of the wage, fringe benefit, material or depreciation from companies under study, and a categorical covariate $Z$ is the type of industry (see Table 1). What the MAR assumption amounts to is that the nonresponse rate depends on the type of industry, but does not vary within a particular industry. Although the MAR assumption may not always hold, it provides an acceptable approximation in many situations [Rubin (1976)] like the SIRD.

Unbiased or approximately unbiased estimators of parameters such as the mean of $Y$ and the mean of $Y$ conditional on $Z$ (e.g., the mean for a particular industry in the SIRD) can be constructed using incomplete $Y$ values, observed $Z$ values and the MAR assumption. In sample surveys, it is often desired to impute nonrespondents and then compute estimates by treating imputed values as observed data and using the standard formulas designed to produce unbiased or approximately unbiased survey estimators in the case of no nonresponse [Kalton and Kasprzyk (1986)].

Let $\{z_1,\ldots,z_s\}$ be the range for $Z$, where $s$ is a fixed positive integer. A simple method that is often applied in practice is to use $z_1,\ldots,z_s$ as "imputation cells" and construct estimators by imputing nonrespondents (or reweighting respondents) within each imputation cell. Although this method of using $z_1,\ldots,z_s$ as imputation cells is simple, it often runs into one or both of the following problems:

1. In practice, imputation cells are not necessarily constructed using all categories according to $Z$ values, because some imputation cells may have small sizes. In some agencies, an internal rule is that in each imputation cell, the number of respondents has to be larger than the number of nonrespondents. Cells with small sizes are often collapsed to achieve this goal. Table 1 displays the nonresponse rate for four variables in the SIRD. Although the overall nonresponse rate (the last line of Table 1) is between 36.8% and 41.1%, nonresponse rates in many industries are higher than 50%. If each imputation cell must have more respondents than nonrespondents, then some $Z$ categories (industries) have to be collapsed. Collapsing cells not only is ad hoc and subjective, but also may violate the MAR assumption and create biased survey estimators. More precisely, let $\tilde{Z}$ be the categorical variable corresponding to the new imputation cells. Then $\tilde{Z}$ is a function of $Z$ and $P(\delta=1|Y,\tilde{Z}) = P(\delta=1|\tilde{Z})$ may not hold.

TABLE 1
*Nonresponse rates (%) for survey items by industry in 2002 SIRD*

| Industry | Wages | Fringe Benefits | Materials | Depreciation |
|---|---|---|---|---|
| Food | 33.8 | 29.4 | 29.5 | 32.7 |
| Beverage and tobacco products | 11.8 | 8.1 | 3.3 | 0.0 |
| Textiles, apparel and leather | 27.3 | 17.0 | 30.6 | 57.0 |
| Paper, printing and support activities | 15.5 | 50.0 | 17.7 | 20.1 |
| Petroleum and coal products | 83.7 | 82.3 | 82.5 | 90.0 |
| Chemicals | 22.4 | 23.8 | 22.8 | 20.1 |
| Plastic and rubber products | 33.5 | 42.5 | 20.4 | 43.5 |
| Nonmetallic mineral products | 17.8 | 24.7 | 22.8 | 21.4 |
| Primary metals | 7.8 | 14.5 | 21.0 | 2.8 |
| Fabricated metal products | 25.2 | 12.4 | 35.3 | 31.6 |
| Machinery | 35.6 | 38.3 | 37.4 | 30.6 |
| Computers and peripheral equipment | 49.7 | 61.5 | 50.6 | 50.1 |
| Communication equipment | 64.5 | 24.7 | 77.4 | 23.0 |
| Semiconducting and other electronic components | 14.7 | 10.6 | 14.4 | 12.5 |
| Navigation, measuring, electronic and control instruments | 56.2 | 62.2 | 56.6 | 58.6 |
| Other computer and electronic products | 36.1 | 37.0 | 38.4 | 21.1 |
| Electrical equipment, appliances and components | 15.8 | 8.2 | 16.8 | 12.3 |
| Motor vehicles and parts | 47.2 | 55.6 | 41.2 | 51.0 |
| Aerospace products and parts | 22.7 | 26.8 | 21.8 | 18.9 |
| Other transportation equipment | 39.3 | 52.5 | 26.3 | 16.4 |
| Furniture and related products | 30.9 | 36.4 | 38.1 | 52.1 |
| Miscellaneous manufacturing | 57.7 | 53.5 | 68.6 | 41.7 |
| Mining, extraction and support activities | 86.6 | 91.8 | 94.4 | 93.1 |
| Construction | 33.2 | 54.3 | 8.7 | 57.3 |
| Trade | 32.2 | 39.4 | 46.0 | 30.6 |
| Publishing | 55.5 | 58.4 | 63.3 | 71.9 |
| Broadcasting and telecommunications | 84.0 | 66.8 | 74.3 | 78.2 |
| Other information | 5.5 | 20.1 | 12.9 | 49.9 |
| Finance, insurance and real estate | 29.8 | 40.9 | 55.5 | 45.3 |
| Architectural, engineering and related services | 32.1 | 26.1 | 38.1 | 67.4 |
| Computer systems design and related services | 43.3 | 40.0 | 39.1 | 45.0 |
| Scientific R&D services | 28.8 | 29.0 | 29.2 | 31.6 |
| Other professional, scientific and technical services | 40.2 | 33.8 | 39.0 | 46.1 |
| Health care services | 12.7 | 12.5 | 14.4 | 35.6 |
| Other nonmanufacturing | 23.2 | 36.8 | 59.1 | 18.0 |
| All industries | 38.1 | 41.1 | 39.3 | 36.8 |

Source: National Science Foundation/Division of Science Resources Statistics, Survey of Industrial Research and Development, 2002.

2. Although unbiasedness of survey estimators is the primary concern in the development of an estimation and/or imputation procedure, the efficiency of survey estimators should also be considered, especially when auxiliary data are available. Let $Z$ be the indicator for several data sets (e.g., data sets from several different years). Suppose that we need to estimate $E(Y|Z = z_1)$. If data from different years ($Z = z_j$, $j > 1$) also carry information about $E(Y|Z = z_1)$, then the estimation efficiency can be improved if we use all data sets, not just the single data set with $Z = z_1$. The question is how to make use of different data sets. Simply pooling different data sets together may introduce some estimation bias, since each data set may have its own population distribution for $Y$.

The purpose of this paper is to study a pseudo empirical likelihood method for estimation and imputation under the MAR assumption and a nonparametric marginal distribution assumption for $Y$ (which is particularly desired for survey data). The empirical likelihood method was developed by Owen (1988) and Qin and Lawless (1994) in the context of independent and identically distributed data and was extended to survey problems (without missing data) by Chen and Qin (1993), Chen and Sitter (1999), Zhong and Rao (2000) and Wu and Rao (2006). When missing data are present, Wang and Rao (2002) and Wang, Linton and Härdle (2004) considered the approach of first imputing missing data based on some method and then applying the empirical likelihood to imputed data to obtain more efficient estimators. Since imputation has to be carried out first, this approach does not deal with the problem of small size imputation cells. Assuming a parametric model on $P(\delta = 1|Y, Z)$ (but allowing the dependence of $Y$ in the response probability), Qin, Leung and Shao (2002) considered estimation with empirical likelihoods putting positive mass to observed $(Y, Z)$ only. Chen and Qin (2006) used a similar approach for a binary $Y$. However, the problem of small size imputation cells was not considered. Furthermore, none of these cited papers contains an imputation procedure using the empirical likelihood approach. We derive empirical likelihood estimators of $E(y)$ and $E(Y|Z = z_j)$ and some imputation procedures that do not involve any ad hoc method of forming imputation cells and provide more efficient estimators than those from the simple approach of using $z_1, \ldots, z_s$ as imputation cells, at the price of assuming a parametric model for $P(Z = z_j|Y)$. To make use of several data sets or utilize all $Z$ categories for imputation and estimation in a small $Z$ category, some model assumption that relates different data sets or categories together is necessary. In survey problems with a continuous variable $Y$, finding a suitable parametric model for $P(Z = z_j|Y)$ is much easier than finding an appropriate parametric model for the conditional distribution of $Y$ given $Z = z_j$.

In our empirical likelihood, there are lots of parameters when $s$ (the number of $Z$ categories) is large, which creates problems in numerical computation of the solution to the likelihood equation. We adopt a pseudo empirical likelihood approach by replacing some nuisance parameters in the likelihood equation with some simple consistent estimators. The resulting estimators may lose some efficiency, but its computation is much more practical. Theoretical properties of the pseudo empirical likelihood estimators are investigated.

Section 2 presents details on the sampling design and model, and results for estimation without imputation. In addition to the derivation of pseudo empirical likelihood estimators, their consistency and asymptotic normality are established. Section 3 considers variance estimation by bootstrapping. In Section 4, we consider several imputation methods related to the results in Section 2. Asymptotic properties of estimators based on imputed data are given. Section 5 examines by simulation the finite sample performance of the proposed estimators, under some response patterns and models. The proofs are sketched in the Appendix.

**2. Pseudo empirical likelihood.** Let $\mathcal{P}$ be a finite population stratified into $H$ strata with $N_h$ units in the $h$th stratum. Assume that $n_h \geq 2$ units are sampled from stratum $h$ according to some probability sampling plan, independently across the strata. When equal probability sampling is used, sampling is either without replacement or with replacement; when unequal probability sampling is applied, we assume that sampling is with replacement, since without replacement unequal probability sampling is not often used because of its complexity and the difficulty in deriving variances of estimators due to the dependence caused by without replacement sampling [Särndal, Swensson and Wretman (1992), Section 3.6]. According to the sampling plan, survey weights $w_{hi}$, $i = 1, \ldots, n_h$, $h = 1, \ldots, H$, are constructed so that for any set of values $\{x_{hi}\}$,

$$E_s\left(\sum_{h=1}^{H}\sum_{i=1}^{n_h} w_{hi} x_{hi}\right) = \frac{1}{N}\sum_{h=1}^{H}\sum_{i=1}^{N_h} x_{hi},$$

where $E_s$ is the expectation with respect to sampling and $N = \sum_{h=1}^{H} N_h$. If $q_{hi}$ is the probability that the $i$th unit in stratum $h$ is in the sample, then the survey weight $w_{hi} = (Nq_{hi})^{-1}$. We consider the asymptotic setting with a fixed $H$, $n_h \to \infty$, and $n_h/N_h \to 0$ for all $h$. This sampling design is commonly used in many business surveys; for example, the Current Employment Survey conducted by the U.S. Bureau of Labor Statistics [Wolter, Shao and Huff (1998)], the Transportation Annual Survey conducted by the U.S. Census Bureau [Census Bureau (1987)] and the Financial Farm Survey conducted by Statistics Canada [Caron (1996)]. In the SIRD discussed in Section 1, strata are created according to industry group and size of companies and, within each stratum, either simple random sampling or probability (proportionate to company size) sampling is used.

Let $Y$ be a variable of interest in the survey and $Z$ be a categorical covariate taking values in $\{z_1, \ldots, z_s\}$. Within stratum $h$, we assume that $(Y, Z)$ is random and follows a superpopulation model with an unknown nonparametric marginal distribution $F_h$ for $Y$ and a parametric probability function

$$P_h(Z = z | Y = y) = f_h(y, z, \beta), \tag{1}$$

where $\beta$ is an unknown parameter vector and $f_h$ is a known function. For each sampled unit, the $Z$ value is always observed, but the $Y$ value may be a nonrespondent. Under the MAR assumption described in Section 1, $P(\delta = 1 | Y, Z) = \phi_h(Z)$ in stratum $h$, where $\phi_h$ is an unknown function. Because of the MAR assumption, we do not need to impose any condition on $\phi_h$ except that $\phi_h(z_j) > 0$ for any $h$ and $z_j$. Without loss of generality, we assume that in stratum $h$, the first $r_h$ sampled units are respondents and the rest of $n_h - r_h$ sampled units are nonrespondents. Thus, the observed data set is

$$\{(Y_{hi}, Z_{hi}), i = 1, \ldots, r_h\} \cup \{Z_{hi}, i = r_h + 1, \ldots, n_h\}, \qquad h = 1, \ldots, H.$$

Let $p_{hi} = dF_h(Y_{hi})$ be the point mass $F_h$ places on $Y_{hi}$. For a particular unit $(h, i)$, if $i \leq r_h$ ($Y_{hi}$ is observed), the likelihood is the joint probability density

$$\phi_h(Z_{hi}) f_h(Y_{hi}, Z_{hi}, \beta) p_{hi};$$

if $i > r_h$ ($Y_{hi}$ is missing), the likelihood is the joint probability density with $Y_{hi}$ integrated out, that is,

$$\int [1 - \phi_h(Z_{hi})] f_h(y, Z_{hi}, \beta) \, dF_h(y) = [1 - \phi_h(Z_{hi})] \int f_h(y, Z_{hi}, \beta) \, dF_h(y).$$

Following the idea in Chen and Sitter (1999), we weight each unit log-likelihood by $w_{hi}$ and obtain the log-likelihood

$$\sum_{h=1}^{H} \left[ \sum_{i=1}^{r_h} w_{hi} \log(\phi_h(Z_{hi}) f_h(Y_{hi}, Z_{hi}, \beta) p_{hi}) + \sum_{i=r_h+1}^{n_h} w_{hi} \log\left( [1 - \phi_h(Z_{hi})] \int f_h(y, Z_{hi}, \beta) \, dF_h(y) \right) \right].$$

Adding the weights $w_{hi}$ is necessary for obtaining approximately unbiased estimators under unequal probability sampling. Since $\phi_h(Z_{hi})$ does not involve $\beta$ and $F_h$, we may focus on

$$L = \sum_{h=1}^{H} \left[ \sum_{i=1}^{r_h} w_{hi} \log(f_h(Y_{hi}, Z_{hi}, \beta) p_{hi}) \right.$$

$$+ \sum_{i=r_h+1}^{n_h} w_{hi} \log\left(\int f_h(y, Z_{hi}, \beta) \, dF_h(y)\right)\Bigg]$$

for the estimation of parameters related to $Y$. Within stratum $h$, let

$$\pi_{hj} = P_h(Z = z_j) = \int f_h(y, z_j, \beta) \, dF_h(y).$$

Since $Z$ takes values $z_1, \ldots, z_s$, $L$ can be written as

$$L = \sum_{h=1}^{H} \left[\sum_{i=1}^{r_h} w_{hi} \log(f_h(Y_{hi}, Z_{hi}, \beta) p_{hi}) + \sum_{j=1}^{s} a_{hj} \log(\pi_{hj})\right],$$

where $a_{hj} = \sum_{i=r_h+1}^{n_h} w_{hi} I_{\{Z_{hi}=z_j\}}$ and $I_A$ is the indicator function of the event $A$. Applying the empirical likelihood approach, we estimate $\beta$ and $F_h$, $h = 1, \ldots, H$, by maximizing $L$ subject to

(2)
$$p_{hi} \geq 0, \qquad \sum_{i=1}^{r_h} p_{hi} = 1, \qquad \sum_{i=1}^{r_h} p_{hi} f_h(Y_{hi}, z_j, \beta) = \pi_{hj},$$

$$j = 1, \ldots, s, h = 1, \ldots, H.$$

Since $F_h$ is nonparametric, its estimate is an empirical distribution with $r_h$ points, the observed $Y_{hi}$, $i = 1, \ldots, r_h$, as the support. Although $Y_{hi}$, $i = 1, \ldots, r_h$, are from the distribution of respondents, we can obtain a valid estimator of the marginal distribution $F_h$ using the covariate information through the terms $\sum_j a_{hj} \log(\pi_{hj})$ in $L$. Using Lagrange multiplier under the constraints in (2) and the usual profile empirical likelihood argument, we can derive that

(3)
$$p_{hi} = \frac{w_{hi}}{\sum_{i=1}^{n_h} w_{hi} - \sum_{j=1}^{s} \frac{a_{hj}}{\pi_{hj}} f_h(Y_{hi}, z_j, \beta)},$$

$$i = 1, \ldots, r_h, h = 1, \ldots, H,$$

and obtain estimators of $\beta$ and $\pi = (\pi_{hj}, j = 1, \ldots, s, h = 1, \ldots, H)$ by maximizing

$$l(\beta, \pi) = \sum_{h=1}^{H} \left[\sum_{i=1}^{r_h} w_{hi} \log\left(\frac{w_{hi} f_h(Y_{hi}, Z_{hi}, \beta)}{\sum_{i=1}^{n_h} w_{hi} - \sum_{j=1}^{s} \frac{a_{hj}}{\pi_{hj}} f_h(Y_{hi}, z_j, \beta)}\right)\right.$$

$$\left. + \sum_{j=1}^{s} a_{hj} \log(\pi_{hj})\right]$$

subject to

(4)
$$\sum_{i=1}^{r_h} \frac{w_{hi}[f_h(Y_{hi}, z_j, \beta) - \pi_{hj}]}{\sum_{i=1}^{n_h} w_{hi} - \sum_{j=1}^{s} \frac{a_{hj}}{\pi_{hj}} f_h(Y_{hi}, z_j, \beta)} = 0,$$

$$j = 1, \ldots, s, h = 1, \ldots, H.$$

When $s$ (the number of $Z$ categories) is not small, it is difficult to maximize $l(\beta, \pi)$ over $(\beta, \pi)$ subject to (4). Numerical solutions may be very computation-intensive to obtain and they may be unreliable. Hence, we apply the idea of pseudo likelihood [Gong and Samaniego (1981)]. Note that consistent estimators of the $\pi_{hj}$ are easy to construct. For example, we may estimate $\pi_{hj}$ by

$$(5) \quad \hat{\pi}_{hj} = \sum_{i=1}^{n_h} w_{hi} I_{\{Z_{hi}=z_j\}} \Big/ \sum_{i=1}^{n_h} w_{hi}.$$

Maximizing the pseudo empirical likelihood $l(\beta, \hat{\pi})$ over $\beta$ results in the maximum pseudo empirical likelihood estimator (MPELE) $\hat{\beta}$, where $\hat{\pi} = (\hat{\pi}_{hj}, j = 1, \ldots, s, h = 1, \ldots, H)$. The MPELE $\hat{\beta}$ can be computed by maximizing $l(\beta, \hat{\pi})$ over $\beta$ using any available software. For example, in the simulation study in Section 5, $\hat{\beta}$ was obtained using FMINSEARCH in MATLAB.

Note that the MPELE is different from the maximum empirical likelihood estimator since $\hat{\pi}$ is not $\pi$. The left-hand side of (4) is not 0 when $\pi$ is replaced by $\hat{\pi}$ and $\beta$ is replaced by the MPELE $\hat{\beta}$, although we show later that it converges to 0 in probability. However, similarly to other cases in which the pseudo likelihood is used, we can directly establish the consistency and asymptotic normality of the MPELE.

Let $\hat{p}_{hi}$ be obtained by using (3) with $\beta$ and $\pi_{hj}$ replaced by $\hat{\beta}$ and $\hat{\pi}_{hj}$, respectively. The distribution function for $Y$ can be estimated by

$$\hat{G}(y) = \sum_{h=1}^{H} W_h \sum_{i=1}^{r_h} \hat{p}_{hi} I_{\{Y_{hi} \leq y\}},$$

where $W_h = N_h/N$, $h = 1, \ldots, H$. However, because the MPELE is used, $\sum_{i=1}^{r_h} \hat{p}_{hi} \neq 1$ (although $\sum_{i=1}^{r_h} \hat{p}_{hi} \to_p 1$) and, hence, $\hat{G}(y)$ is not a distribution function. A modified distribution estimator for $Y$ is

$$\hat{F}(y) = \sum_{h=1}^{H} W_h \sum_{i=1}^{r_h} \hat{p}_{hi} I_{\{Y_{hi} \leq y\}} \Big/ \sum_{h=1}^{H} W_h \sum_{i=1}^{r_h} \hat{p}_{hi}$$

$$= \sum_{h=1}^{H} \sum_{i=1}^{r_h} \tilde{p}_{hi} I_{\{Y_{hi} \leq y\}} \Big/ \sum_{h=1}^{H} \sum_{i=1}^{r_h} \tilde{p}_{hi},$$

where $\tilde{p}_{hi} = W_h \hat{p}_{hi}$. If the parameter of interest is the finite population mean $\bar{Y} = \sum_{h=1}^{H} \sum_{i=1}^{N_h} Y_{hi}/N$, its MPELE is

$$(6) \quad \hat{\bar{Y}} = \int y \, d\hat{F}(y) = \sum_{h=1}^{H} \sum_{i=1}^{r_h} \tilde{p}_{hi} Y_{hi} \Big/ \sum_{h=1}^{H} \sum_{i=1}^{r_h} \tilde{p}_{hi}.$$

Given $Z = z_j$, the conditional distribution of $Y$ can be estimated by

$$\hat{F}_j(y) = \sum_{h=1}^{H}\sum_{i=1}^{r_h} \tilde{p}_{hi} f_h(Y_{hi}, z_j, \hat{\beta}) I_{\{Y_{hi} \leq y\}} \Big/ \sum_{h=1}^{H}\sum_{i=1}^{r_h} \tilde{p}_{hi} f_h(Y_{hi}, z_j, \hat{\beta}).$$

If the parameter of interest is the cell mean $\bar{Y}_j$, the finite population mean of $Y$ given $Z = z_j$, its MPELE is

(7)
$$\begin{aligned}
\hat{\bar{Y}}_j &= \int y \, d\hat{F}_j(y) \\
&= \sum_{h=1}^{H}\sum_{i=1}^{r_h} \tilde{p}_{hi} f_h(Y_{hi}, z_j, \hat{\beta}) Y_{hi} \Big/ \sum_{h=1}^{H}\sum_{i=1}^{r_h} \tilde{p}_{hi} f_h(Y_{hi}, z_j, \hat{\beta}).
\end{aligned}$$

The following result shows that the MPELE $\hat{\beta}$, $\hat{\bar{Y}}$ and $\hat{\bar{Y}}_j$ are consistent estimators and are asymptotically normal. The proofs are given in the Appendix.

THEOREM 1. *Assume MAR as described in Section 1 and model (1). Suppose that regularity conditions* (i)–(v) *stated in the Appendix. Then, there exists a sequence* $\{\hat{\beta}_n, n = 1, 2, \ldots\}$ *such that* $\|\hat{\beta}_n - \beta_0\| \leq n^{-1/3}$ *and as* $n \to \infty$,

(8) $$P\left(\frac{\partial l(\hat{\beta}_n, \hat{\pi})}{\partial \beta} = 0\right) \to 1 \quad \text{and} \quad \sqrt{n}(\hat{\beta}_n - \beta_0) \to_d N(0, \Lambda),$$

*where* $\Lambda$ *is a positive definite matrix. Furthermore, if condition* (iv) *in the Appendix holds, then*

(9) $$\sqrt{n}(\hat{\bar{Y}} - \bar{Y}) \to_d N(0, \sigma^2) \quad \text{and} \quad \sqrt{n}(\hat{\bar{Y}}_j - \bar{Y}_j) \to_d N(0, \sigma_j^2),$$
$$j = 1, \ldots, s,$$

*where* $\sigma^2$ *[given by (20)–(21) in the Appendix] and* $\sigma_j^2$ *are some positive constants.*

The simple method of reweighting respondents within each imputation cell (see Section 1) produces the following estimators of $\bar{Y}$ and $\bar{Y}_j$:

(10) $$\tilde{\bar{Y}} = \sum_{j=1}^{s}\sum_{h=1}^{H}\sum_{i=1}^{n_h} w_{hi} I_{\{Z_{hi}=z_j\}} \tilde{\bar{Y}}_{hj} \Big/ \sum_{h=1}^{H}\sum_{i=1}^{n_h} w_{hi}$$

and

(11) $$\tilde{\bar{Y}}_j = \sum_{h=1}^{H}\sum_{i=1}^{n_h} w_{hi} I_{\{Z_{hi}=z_j\}} \tilde{\bar{Y}}_{hj} \Big/ \sum_{h=1}^{H}\sum_{i=1}^{n_h} w_{hi} I_{\{Z_{hi}=z_j\}},$$

where

$$\tag{12} \tilde{\hat{Y}}_{hj} = \sum_{i=1}^{r_h} w_{hi} Y_{hi} I_{\{Z_{hi}=z_j\}} \Big/ \sum_{i=1}^{r_h} w_{hi} I_{\{Z_{hi}=z_j\}}.$$

Some comparisons of these estimators with the MPELE are made in a simulation study (Section 5).

**3. Variance estimation by bootstrapping.** It is a common practice in sample surveys to report a variance estimate for each estimate of the parameter of interest. We focus on the most commonly used estimators, the mean estimator $\hat{\bar{Y}}$ in (6) and the cell mean estimator $\hat{\bar{Y}}_j$ in (7). Because both $\hat{\bar{Y}}$ and $\hat{\bar{Y}}_j$ are complex functions of $\hat{\beta}$ and $\hat{p}_{hi}$, it is difficult to derive an analytic form of their asymptotic variances, $\sigma^2$ and $\sigma_j^2$ in (9). It is shown in the Appendix that $\sigma^2$ is equal to the limit of $\sum_{h=1}^{H} n N_h^2 \sigma_h^2 / (n_h N^2)$, where $\sigma_h^2$ is given in (21) in the Appendix and has a complicate form. Thus, we apply the bootstrap method, which consists of the following steps. In the following $\hat{\theta}$ denotes $\hat{\beta}$, $\hat{\bar{Y}}$ or $\hat{\bar{Y}}_j$.

1. Within stratum $h$, draw a simple random sample of size $n_h$ with replacement from the set of sampled units (respondents or nonrespondents). Carry out this procedure independently across strata. For each unit in the bootstrap sample, the bootstrap data are the $Z$ and $Y$ values (if the $Y$ is missing, the bootstrap datum is treated as missing) and its survey weight.
2. Compute $\hat{\theta}^*$, which is the same as $\hat{\theta}$ but with the original data replaced by the bootstrap data generated in step 1.
3. Repeat the previous steps independently $B$ times and obtain $\hat{\theta}^{*1}, \ldots, \hat{\theta}^{*B}$. Estimate the variance of $\hat{\theta}$ by the sample variance of $\hat{\theta}^{*1}, \ldots, \hat{\theta}^{*B}$.

If there is no nonresponse, then the previously described bootstrap produces consistent variance estimators for mean estimators [see, e.g., Shao and Tu (1995), Chapter 6]. However, no theory is available for the bootstrap when empirical likelihoods are used for nonrespondents. We establish the following result for the asymptotic validity of the bootstrap.

THEOREM 2. *Assume the conditions in Theorem 1. Let $l^*(\cdot, \cdot)$ be the bootstrap analog of $l(\cdot, \cdot)$ and $\hat{\pi}^* = (\hat{\pi}^*_{hj}, j = 1, \ldots, s, h = 1, \ldots, H)$ with $\hat{\pi}^*_{hj}$ is the bootstrap analog of $\hat{\pi}_{hj}$ in (5). Then, there exists a sequence $\{\hat{\beta}^*, n = 1, 2, \ldots\}$ such that $\|\hat{\beta}^* - \hat{\beta}\| \leq n^{-1/3}$ and as $n \to \infty$,*

$$\tag{13} P_*\left(\frac{\partial l^*(\hat{\beta}^*, \hat{\pi}^*)}{\partial \beta} = 0\right) \to_p 1 \quad \text{and} \quad \sqrt{n}(\hat{\beta}^* - \hat{\beta}) \to_{d^*} N(0, \Lambda),$$

where $\Lambda$ is given in (8), $P_*$ denotes the bootstrap probability conditional on the data, and $\vartheta_n^* \to_{d^*} \vartheta$ means $P_*(\vartheta_n^* \in B) - P(\vartheta \in B) \to_p 0$ for any Borel set $B$. Furthermore, if condition (iv) of Theorem 1 also holds, then

$$\sqrt{n}(\hat{\bar{Y}}^* - \hat{\bar{Y}}) \to_{d^*} N(0, \sigma^2) \quad \text{and} \quad \sqrt{n}(\hat{\bar{Y}}_j^* - \hat{\bar{Y}}_j) \to_{d^*} N(0, \sigma_j^2),$$

where $\sigma^2$ and $\sigma_j^2$ are defined in (9).

**4. Imputation.** Imputation is often carried out for practical reasons [Kalton and Kasprzyk (1986)]. After imputation, estimates of parameters are computed by treating imputed values as observed data and using the standard formulas for the case of no nonresponse. In this section we consider imputation for the estimation of the population mean $\bar{Y}$ and the population cell mean $\bar{Y}_j$. Let $\hat{Y}_{hi} = Y_{hi}$ if $Y_{hi}$ is a respondent and let $\hat{Y}_{hi}$ be an imputed value if $Y_{hi}$ is a nonrespondent. After imputation, the population mean $\bar{Y}$ and cell mean $\bar{Y}_j$ are estimated by

$$\hat{\bar{Y}}_I = \sum_{h=1}^{H} \sum_{i=1}^{n_h} w_{hi} \hat{Y}_{hi} \tag{14}$$

and

$$\hat{\bar{Y}}_{jI} = \sum_{h=1}^{H} \sum_{i=1}^{n_h} w_{hi} \hat{Y}_{hi} I_{\{Z_{hi}=z_j\}} \Big/ \sum_{h=1}^{H} \sum_{i=1}^{n_h} w_{hi} I_{\{Z_{hi}=z_j\}}, \tag{15}$$

respectively. The simple method of using $z_1, \ldots, z_s$ as imputation cells imputes nonrespondents in an imputation cell using respondents in the same cell only. The simple mean imputation method imputes each nonrespondent in stratum $h$ with $Z = z_j$ by the cell sample mean $\tilde{\bar{Y}}_{hj}$ given in (12). The simple random imputation method imputes each nonrespondent in stratum $h$ with $Z = z_j$ by a random sample with replacement from respondents in stratum $h$ with $Z = z_j$, where each $Y_{hi}$ with $Z_{hi} = z_j$ has probability $w_{hi} I_{\{Z_{hi}=z_j\}} / \sum_{i=1}^{r_h} w_{hi} I_{\{Z_{hi}=z_j\}}$ to be selected, $i = 1, \ldots, r_h$. Problems of these simple imputation methods are discussed in Section 1.

Using the MPELE estimators developed in Section 2, we consider the following two imputation procedures:

1. Pseudo Likelihood Mean Imputation. For each nonrespondent in stratum $h$ with $Z = z_j$, its imputed $Y$ value is the mean estimator

$$\hat{\bar{Y}}_{hj} = \sum_{i=1}^{r_h} \hat{p}_{hi} f_h(Y_{hi}, z_j, \hat{\beta}) Y_{hi} \Big/ \sum_{i=1}^{r_h} \hat{p}_{hi} f_h(Y_{hi}, z_j, \hat{\beta}).$$

2. Pseudo Likelihood Random Imputation. Each nonrespondent in stratum $h$ with $Z = z_j$ is imputed by a random sample with replacement from all respondents in stratum $h$, where the probability of each $Y_{hi}$ to be selected is $f_h(Y_{hi}, z_j, \hat{\beta}) \hat{p}_{hi} / \sum_{i=1}^{r_h} f_h(Y_{hi}, z_j, \hat{\beta}) \hat{p}_{hi}$, $i = 1, \ldots, r_h$.

The following result shows that the estimators of $\bar{Y}$ and $\bar{Y}_j$ based on these two imputation procedures are consistent and asymptotically normal.

THEOREM 3. *Under the conditions of Theorem 1, for either pseudo likelihood mean imputation or pseudo likelihood random imputation,*

$$\sqrt{n}(\hat{\bar{Y}}_I - \bar{Y}) \to_d N(0, \sigma_I^2) \quad and$$
$$\sqrt{n}(\hat{\bar{Y}}_{jI} - \bar{Y}_j) \to_d N(0, \sigma_{jI}^2), \qquad j = 1, \ldots, s,$$

*where $\sigma_I^2$ and $\sigma_{jI}^2$ are some positive constants.*

The main difference between the simple (mean or random) imputation method and the pseudo likelihood (mean or random) imputation method is that the former restricts imputation within each $Z$ category whereas the latter uses all respondents with appropriate weighting. Hence, the latter avoids the problems described in Section 1 for simple imputation and is more efficient when our assumption on $P(Z = z_j | Y)$ holds.

The asymptotic variances $\sigma_I^2$ and $\sigma_{jI}^2$ do not have simple analytic forms. Variance estimation can be carried out using the bootstrap procedure described in Section 3. It should be emphasized that, to address the variability caused by imputation, nonrespondents in each bootstrap data set must be imputed using the bootstrap data and the same imputation method as that used to impute the original data set, as suggested by Shao and Sitter (1996).

**5. Simulation results.** In this section, we evaluate by simulation the finite sample properties of the MPELE and the pseudo likelihood imputation. We create a finite population similar to the Current Establishment Survey conducted by the U.S. Bureau of Labor Statistics. We choose four different industries as four strata with sizes $N_1 = 3370$, $N_2 = 2910$, $N_3 = 5430$ and $N_4 = 4110$. The variable $Y$ is the total pay for each establishment and values of $Y$ in stratum $h$ are generated from a superpopulation $F_h$. The form of $F_h$ is chosen to be the gamma distribution and $F_1 = \Gamma(43, 0.20)$, $F_2 = \Gamma(42, 0.19)$, $F_3 = \Gamma(38, 0.20)$ and $F_4 = \Gamma(50, 0.17)$, where $\Gamma(a, b)$ denotes the gamma distribution with shape parameter $a$ and scale parameter $b$. The parameters in $F_h$'s are chosen to match the mean and variance of a real data set from the Current Establishment Survey. The covariate $Z \in \{1, 2, 3, 4, 5\}$ is generated by the proportional-odds model

$$\log \frac{P(Z \leq j | Y = y)}{P(Z > j | Y = y)} = j + \beta y, \qquad j = 1, 2, 3, 4,$$

where $\beta$ is an unknown parameter whose value in the simulation is $-0.4$.

The sampling plan is stratified simple random sampling without replacement. In each stratum, the sampling fraction is 0.03. For each sampled unit, the $Y$ respondent is generated according to the response probability function

$$P(\delta = 1|Z = j) = \frac{\exp\{-0.1 + \gamma j\}}{1 + \exp\{-0.1 + \gamma j\}}$$

with an unknown parameter $\gamma$. The following table lists values of $\gamma$ considered in the simulation, the response rate for each $Z$, and the mean response rate $E[P(\delta = 1|Z)]$:

| $\gamma$ | 0.7 | 0.5 | 0.3 | 0.1 | $-0.1$ |
|---|---|---|---|---|---|
| $P(\delta = 1|Z = 1)$ | 0.6457 | 0.5978 | 0.5498 | 0.5000 | 0.4502 |
| $P(\delta = 1|Z = 2)$ | 0.7858 | 0.7109 | 0.6225 | 0.5250 | 0.4256 |
| $P(\delta = 1|Z = 3)$ | 0.8808 | 0.8022 | 0.6900 | 0.5498 | 0.4013 |
| $P(\delta = 1|Z = 4)$ | 0.9307 | 0.8699 | 0.7503 | 0.5744 | 0.3775 |
| $P(\delta = 1|Z = 5)$ | 0.9677 | 0.9168 | 0.8022 | 0.5987 | 0.3542 |
| $E[P(\delta = 1|Z)]$ | 0.8852 | 0.8224 | 0.7154 | 0.5634 | 0.3888 |

For each $\gamma$, we run the simulation for 1000 times. Table 2 reports the variance (Var) of the proposed MPELE estimators $\hat{\beta}$, $\hat{\bar{Y}}$ in (6), $\hat{\bar{Y}}_j$ in (7), $\hat{\bar{Y}}_I$ in (14) and $\hat{\bar{Y}}_{jI}$ in (15), based on either pseudo likelihood mean imputation or pseudo likelihood random imputation. All the relative biases are less than 0.3% and hence not reported. To compare the efficiency of the MPELE estimators of the means (with imputation or without imputation) with the simple estimators using the $Z$ categories as imputation cells, we report the ratios of mean square errors (Rat) in Table 2. Each MPELE is compared with its counterpart; that is, $\hat{\bar{Y}}$ in (6) is compared with $\tilde{\bar{Y}}$ in (10), $\hat{\bar{Y}}_j$ in (7) is compared with $\tilde{\bar{Y}}_j$ in (11), and $\hat{\bar{Y}}_I$ in (14) [or $\hat{\bar{Y}}_{jI}$ in (15)] with pseudo likelihood mean (or random) imputation is compared with $\hat{\bar{Y}}_I$ in (14) [or $\hat{\bar{Y}}_{jI}$ in (15)] with simple mean (or random) imputation described in Section 4. To study the performance of the bootstrap, Table 2 also reports the bootstrap variance estimators (Vboot) with $B = 200$ for the MPELE estimators and estimators based on pseudo likelihood mean or random imputation. In addition, Table 2 reports the simulation coverage probabilities (CP) of confidence intervals of the form

$$\text{point estimate} \pm 1.96\sqrt{\text{Vboot}}$$

which approximately have nominal coverage probability 95%.

The results in Table 2 can be summarized as follows:

1. In all cases, the proposed estimators based on the pseudo empirical likelihood (with imputation or not) perform well in terms of the relative bias (less than 0.3%) and variance. For the cell mean estimation, our proposed estimators are much more efficient than the simple estimators based on imputation cells. The ratio of the MSEs can be as small as 0.112 and is always less than 0.5 for estimators without imputation. For the overall mean estimation, our proposed estimators are still more efficient but the improvement is very little. This is expected since "borrowing strength" from other imputation cells has a larger impact for the cell mean estimation than for the overall mean estimation.
2. When the response probability decreases, the variances of our proposed cell mean estimators increase, but their relative efficiencies to the simple estimators increase a great deal especially for the imputation methods. Note that the number of observations within a $Z$ category increases as $Z$ value increases, which results in a decrease in the gain in efficiency from our proposed cell mean estimators. These observations from Table 2 indicate that our proposed cell mean estimators can improve the efficiency over the simple estimators particularly in the cases where some imputation cells have relatively small sample sizes and/or larger number of nonrespondents.
3. In terms of the efficiency, the estimator without imputation is ranked the first, and the estimator with random imputation is ranked the last. However, imputation may be carried out because of practical reasons other than efficiency.
4. The bootstrap variance estimator works well in most cases in terms of its bias and variance (simulation variances are less than $10^{-3}$ and not reported). The coverage probabilities of the confidence intervals are all around 95% with the worst case 91.7% ($\gamma = 0.3$, MPELE).

## APPENDIX

REGULARITY CONDITIONS FOR THEOREM 1.

(i) For each $h$, there are positive constants $k_h$ and $c_h$ such that $\frac{n_h}{n} = k_h + o(1)$ and $\frac{N_h}{N} = c_h + o(n^{-1/2})$, where $n = \sum_{h=1}^{H} n_h$.

(ii) $\max_{i \leq N_h} n_h/(Nq_{hi}) = O(1)$ and $N_h^{-2} \sum_{i=1}^{N_h} n_h/q_{hi} \to d_h$ for some constant $d_h$, $h = 1, \ldots, H$.

(iii) $f_h(y, z, \beta)$ is twice continuously differentiable in $\beta$ for any $h$, $y$ and $z$, and functions $\|\frac{\partial 2 \log f_h(y,z,\beta)}{\partial \beta \partial \beta^\tau}\|^2$, $\|\frac{\partial f_h(y,z_j,\beta)}{\partial \beta}\|^3$, $\|\frac{\partial^2 f_h(y,z_j,\beta)}{\partial \beta \partial \beta^\tau}\|^2$ and $\|[\frac{\partial f_h(y,z_j,\beta)}{\partial \beta}] \times [\frac{\partial f_h(y,z_k,\beta)}{\partial \beta}]^\tau\|^2$ are bounded by a function $g(Y)$ with $E_h[g(Y)] < \infty$ in a neighborhood of $\beta_0$, the true value of $\beta$, $j, k = 1, \ldots, s$, where $E_h$ is the expectation under $F_h$.

TABLE 2
*Simulation results based on 1000 simulation rounds and 200 bootstrap rounds*

| Method | | Response pattern | | | | | | | | | | | | | | | | | | | |
|---|---|---|---|---|---|---|---|---|---|---|---|---|---|---|---|---|---|---|---|---|---|
| | | $\gamma = 0.7$ | | | | $\gamma = 0.5$ | | | | $\gamma = 0.3$ | | | | $\gamma = 0.1$ | | | | $\gamma = -0.1$ | | | |
| | | Var | Vboot | CP | Rat | Var | Vboot | CP | Rat | Var | Vboot | CP | Rat | Var | Vboot | CP | Rat | Var | Vboot | CP | Rat |
| MPELE | $\beta$ | 0.0001 | | | | 0.0001 | | | | 0.0001 | | | | 0.0001 | | | | 0.0001 | | | |
| | $Y$ | 0.0036 | 0.0036 | 96.9 | 0.971 | 0.0038 | 0.0038 | 95.2 | 0.974 | 0.0042 | 0.0044 | 93.8 | 0.972 | 0.0054 | 0.0055 | 93.7 | 0.956 | 0.0079 | 0.0082 | 94.9 | 0.940 |
| | $Y_1$ | 0.0051 | 0.0045 | 92.4 | 0.114 | 0.0056 | 0.0049 | 93.2 | 0.112 | 0.0060 | 0.0056 | 92.7 | 0.115 | 0.0076 | 0.0069 | 92.8 | 0.120 | 0.0102 | 0.0096 | 93.4 | 0.143 |
| | $Y_2$ | 0.0045 | 0.0040 | 94.3 | 0.141 | 0.0049 | 0.0042 | 93.5 | 0.141 | 0.0053 | 0.0049 | 91.7 | 0.149 | 0.0065 | 0.0061 | 93.4 | 0.147 | 0.0091 | 0.0086 | 92.9 | 0.166 |
| | $Y_3$ | 0.0038 | 0.0037 | 94.2 | 0.239 | 0.0042 | 0.0039 | 94.3 | 0.239 | 0.0046 | 0.0045 | 92.9 | 0.241 | 0.0057 | 0.0056 | 94.0 | 0.226 | 0.0083 | 0.0081 | 93.2 | 0.252 |
| | $Y_4$ | 0.0038 | 0.0036 | 95.8 | 0.244 | 0.0039 | 0.0038 | 94.5 | 0.263 | 0.0044 | 0.0044 | 93.7 | 0.249 | 0.0058 | 0.0056 | 92.3 | 0.248 | 0.0084 | 0.0082 | 94.7 | 0.225 |
| | $Y_5$ | 0.0048 | 0.0046 | 95.7 | 0.410 | 0.0050 | 0.0049 | 95.2 | 0.424 | 0.0056 | 0.0056 | 94.0 | 0.461 | 0.0073 | 0.0072 | 93.5 | 0.423 | 0.0112 | 0.0112 | 92.9 | 0.377 |
| Mean imputation | $Y$ | 0.0036 | 0.0036 | 96.9 | 0.970 | 0.0038 | 0.0038 | 95.2 | 0.973 | 0.0042 | 0.0043 | 93.4 | 0.967 | 0.0054 | 0.0055 | 93.7 | 0.951 | 0.0079 | 0.0080 | 95.3 | 0.938 |
| | $Y_1$ | 0.0231 | 0.0229 | 95.9 | 0.493 | 0.0216 | 0.0223 | 95.3 | 0.443 | 0.0213 | 0.0213 | 94.3 | 0.396 | 0.0215 | 0.0216 | 92.7 | 0.352 | 0.0225 | 0.0230 | 94.1 | 0.341 |
| | $Y_2$ | 0.0204 | 0.0204 | 94.2 | 0.683 | 0.0191 | 0.0193 | 94.9 | 0.590 | 0.0180 | 0.0184 | 93.8 | 0.481 | 0.0186 | 0.0182 | 95.0 | 0.389 | 0.0188 | 0.0184 | 95.5 | 0.329 |
| | $Y_3$ | 0.0143 | 0.0142 | 95.7 | 0.831 | 0.0130 | 0.0134 | 96.1 | 0.741 | 0.0124 | 0.0129 | 93.6 | 0.613 | 0.0114 | 0.0124 | 94.6 | 0.465 | 0.0132 | 0.0134 | 94.5 | 0.381 |
| | $Y_4$ | 0.0134 | 0.0136 | 95.7 | 0.904 | 0.0137 | 0.0133 | 95.1 | 0.827 | 0.0121 | 0.0125 | 92.8 | 0.690 | 0.0124 | 0.0122 | 94.0 | 0.505 | 0.0128 | 0.0132 | 95.6 | 0.336 |
| | $Y_5$ | 0.0106 | 0.0109 | 95.5 | 0.966 | 0.0109 | 0.0108 | 94.6 | 0.913 | 0.0106 | 0.0109 | 95.2 | 0.812 | 0.0108 | 0.0113 | 93.4 | 0.660 | 0.0139 | 0.0141 | 94.1 | 0.470 |
| Random imputation | $Y$ | 0.0038 | 0.0039 | 97.0 | 0.987 | 0.0043 | 0.0043 | 94.5 | 1.003 | 0.0050 | 0.0052 | 95.3 | 0.893 | 0.0066 | 0.0067 | 93.4 | 0.964 | 0.0097 | 0.0098 | 95.6 | 0.970 |
| | $Y_1$ | 0.0315 | 0.0322 | 95.4 | 0.559 | 0.0324 | 0.0329 | 95.7 | 0.534 | 0.0322 | 0.0334 | 95.4 | 0.497 | 0.0363 | 0.0348 | 94.8 | 0.498 | 0.0364 | 0.0373 | 94.6 | 0.446 |
| | $Y_2$ | 0.0245 | 0.0249 | 94.5 | 0.736 | 0.0254 | 0.0255 | 93.9 | 0.682 | 0.0267 | 0.0264 | 95.3 | 0.541 | 0.0283 | 0.0281 | 94.0 | 0.507 | 0.0311 | 0.0302 | 95.5 | 0.460 |
| | $Y_3$ | 0.0159 | 0.0157 | 96.1 | 0.802 | 0.0156 | 0.0160 | 95.5 | 0.766 | 0.0164 | 0.0171 | 95.2 | 0.651 | 0.0176 | 0.0184 | 94.4 | 0.567 | 0.0218 | 0.0215 | 95.0 | 0.517 |
| | $Y_4$ | 0.0142 | 0.0145 | 95.8 | 0.920 | 0.0154 | 0.0150 | 94.8 | 0.805 | 0.0155 | 0.0159 | 93.4 | 0.744 | 0.0182 | 0.0175 | 94.3 | 0.627 | 0.0205 | 0.0210 | 95.3 | 0.438 |
| | $Y_5$ | 0.0109 | 0.0112 | 95.2 | 0.971 | 0.0118 | 0.0116 | 94.9 | 0.901 | 0.0124 | 0.0128 | 95.2 | 0.817 | 0.0148 | 0.0153 | 94.6 | 0.677 | 0.0204 | 0.0204 | 94.8 | 0.541 |

The relative biases are all less than 0.3%. Var: variance of the estimators. Vboot: bootstrap variance estimator. CP: coverage probability in % of 95% confidence interval. Rat: ratio of the MSE to the MSE of simple estimator.

(iv) For each $h$, there exists a $z_j$ such that $E_h[\frac{\partial f_h(Y,z_j,\beta_0)}{\partial \beta}][\frac{\partial f_h(Y,z_j,\beta_0)}{\partial \beta}]^\tau$ is positive definite.

(v) $\phi_h(z_j)$ is positive for any $h$ and $z_j$.

(vi) $\|y\frac{\partial f_h(y,z_j,\beta)}{\partial \beta}\|^2$ is bounded by an integrable function in a neighborhood of $\beta_0$ for each $j$.

PROOF OF THEOREM 1. For any function $f(\beta)$, we use the notation $f'(\beta) = \partial f(\beta)/\partial \beta$ and $f''(\beta) = \partial 2 f(\beta)/\partial \beta \partial \beta^\tau$. Let $B_n = \{\beta : \|\beta - \beta_0\| \leq n^{-1/3}\}$, $\partial B_n = \{\beta : \|\beta - \beta_0\| = n^{-1/3}\}$, and $l(\beta) = l(\beta, \hat{\pi})$. For the first conclusion in (8), it suffices to show that

$$(16) \qquad P(l(\beta) - l(\beta_0) < 0 \text{ for all } \beta \in \partial B_n) \to 1.$$

The function $l(\beta)$ is equal to $\sum_{h=1}^H l_h(\beta)$ plus a term that does not depend on $\beta$, where

$$l_h(\beta) = \sum_{i=1}^{r_h} w_{hi} \log f_h(Y_{hi}, Z_{hi}, \beta)$$
$$- \sum_{i=1}^{r_h} w_{hi} \log\left(1 - \sum_{j=1}^{s} \frac{a_{hj}}{\hat{W}_h \hat{\pi}_{hj}} f_h(Y_{hi}, z_j, \beta)\right),$$

and $\hat{W}_h = \sum_{i=1}^{n_h} w_{hi} / \sum_{h=1}^{H} \sum_{i=1}^{n_h} w_{hi}$. Thus, it suffices to show that (16) holds with $l(\beta)$ replaced by $l_h(\beta)$ for each $h$. When $\beta \in \partial B_n$,

$$(17) \quad l_h(\beta) - l_h(\beta_0) = (\beta - \beta_0)^\tau l_h'(\beta_0) + \tfrac{1}{2}(\beta - \beta_0)^\tau l_h''(\beta^\star)(\beta - \beta_0),$$

where $\beta^\star$ is between $\beta$ and $\beta_0$. Define $A_n = l_h'(\beta_0)/W_h$, $c_{n,j} = \frac{a_{hj}}{\hat{W}_h \hat{\pi}_{hj}} - [1 - \phi_h(z_j)]$, and

$$d_{n,i} = \frac{[\sum_{j=1}^s f_h(Y_{hi}, z_j, \beta_0) c_{n,j} / \sum_{j=1}^s \phi_h(z_j) f_h(Y_{hi}, z_j, \beta_0)]^2}{1 - \sum_{j=1}^s f_h(Y_{hi}, z_j, \beta_0) c_{n,j} / \sum_{j=1}^s \phi_h(z_j) f_h(Y_{hi}, z_j, \beta_0)}.$$

Then $A_n = A_{1n} + A_{2n} + A_{3n} + A_{4n}$, where

$$A_{1n} = \frac{1}{W_h} \sum_{i=1}^{r_h} w_{hi} \frac{f_h'(Y_{hi}, Z_{hi}, \beta_0)}{f_h(Y_{hi}, Z_{hi}, \beta_0)},$$

$$A_{2n} = \sum_{j=1}^{s} \frac{a_{hj}}{\hat{W}_h \hat{\pi}_{hj}} \frac{1}{W_h} \sum_{i=1}^{r_h} w_{hi} \frac{f_h'(Y_{hi}, z_j, \beta_0)}{\sum_{j=1}^s \phi_h(z_j) f_h(Y_{hi}, z_j, \beta_0)},$$

$$A_{3n} = \sum_{j=1}^{s} \frac{a_{hj}}{\hat{W}_h \hat{\pi}_{hj}} \sum_{j'=1}^{s} c_{n,j'} \frac{1}{W_h} \sum_{i=1}^{r_h} w_{hi} \frac{f_h'(Y_{hi}, z_j, \beta_0) f_h(Y_{hi}, z_{j'}, \beta_0)}{[\sum_{j=1}^s \phi_h(z_j) f_h(Y_{hi}, z_j, \beta_0)]^2},$$

$$A_{4n} = \sum_{j=1}^{s} \frac{a_{hj}}{\hat{W}_h \hat{\pi}_{hj}} \frac{1}{W_h} \sum_{i=1}^{r_h} w_{hi} \frac{f_h'(Y_{hi}, z_j, \beta_0)}{\sum_{j=1}^s \phi_h(z_j) f_h(Y_{hi}, z_j, \beta_0)} d_{n,i}.$$

We finish the proof of (16) by the following four steps.

**Step 1.** Show that $A_n \to_p 0$. Let $E_c$ denote the conditional expectation of $(Y, Z)$ given $\delta = 1$ in stratum $h$. Then $A_{1n} \to_p P_h(\delta = 1)E_c \frac{f'_h(y,z,\beta_0)}{f_h(y,z,\beta_0)} = \sum_{j=1}^{s} \phi_h(z_j)\pi'_{hj}(\beta_0)$, where $\pi_{hj}(\beta) = \int f_h(y, z_j, \beta)\, dF_h(y)$. Since $\phi_h(z_j)$ is positive, there exists a positive constant $\omega_0$ such that $\phi_h(z_j) \geq \omega_0$ for $h = 1, \ldots, H$ and $j = 1, \ldots, s$. Since $\omega_0 \leq \sum_{j=1}^{s}\phi_h(z_j)f_h(Y_{hi}, z_j, \beta_0) \leq 1$ and $|\sum_{j=1}^{s} f_h(Y_{hi}, z_j, \beta_0)c_{n,j}| \leq \sum_{j=1}^{s}|c_{n,j}|$, we have $d_{n,i} = o_p(\sum_{j=1}^{s}|c_{n,j}|)$ uniformly in $i$. Also, $\sqrt{n_h}c_{n,j} \to_d N(0, \tau^2)$ for some $\tau^2$. Hence,

$$A_{4n} = o_p\left(\sum_{j=1}^{s}|c_{n,j}|\sum_{j=1}^{s}\frac{|a_{hj}|}{\hat{W}_h\hat{\pi}_{hj}}\frac{1}{W_h}\sum_{i=1}^{r_h}w_{hi}\frac{\|f'_h(Y_{hi}, z_j, \beta_0)\|}{\sum_{j=1}^{s}\phi_h(z_j)f_h(Y_{hi}, z_j, \beta_0)}\right)$$
$$= o_p(n_h^{-1/2}).$$

Similarly, $A_{3n} = O_p(n_h^{-1/2}) = o_p(1)$. Finally, $A_{2n} \to_p \sum_{j=1}^{s}[1 - \phi_h(z_j)]\pi'_{hj}(\beta_0)$. Then $A_n \to_p 0$ follows from $\sum_{j=1}^{s}\phi_h(z_j)\pi_{hj}(\beta) + \sum_{j=1}^{s}[1 - \phi_h(z_j)]\pi_{hj}(\beta) = 1$.

**Step 2.** Show that $\sqrt{n_h}A_n \to_d N(0, \Sigma_h)$, where $\Sigma_h$ is $p \times p$ and $p$ is the dimension of $\beta$. Write $A_{1n} + A_{2n} + A_{3n}$ as $g(\frac{1}{W_h}\sum_{i=1}^{n_h}w_{hi}\phi(x_{hi}, \beta_0))$, where $x_{hi} = (\delta_{hi}, Y_{hi}, Z_{hi})$,

$$\phi(x_{hi}, \beta_0) = \left\{\delta_{hi}\frac{f'_h(Y_{hi}, Z_{hi}, \beta_0)}{f_h(Y_{hi}, Z_{hi}, \beta_0)}, (1-\delta_{hi})I\{Z_{hi} = z_j\}|_{j=1,\ldots,s},\right.$$

$$I\{Z_{hi} = z_j\}|_{j=1,\ldots,s}, \frac{\delta_{hi}f'_h(Y_{hi}, z_j, \beta_0)}{\sum_{j=1}^{s}\phi_h(z_j)f_h(Y_{hi}, z_j, \beta_0)}\bigg|_{j=1,\ldots,s},$$

$$\left.\frac{\delta_{hi}f'_h(Y_{hi}, z_j, \beta_0)f_h(Y_{hi}, z_{j'}, \beta_0)}{[\sum_{j=1}^{s}\phi_h(z_j)f_h(Y_{hi}, z_j, \beta_0)]^2}\bigg|_{j,j'=1,\ldots,s}\right\},$$

and $g$ is defined as

$$g(\zeta, \eta_1, \ldots, \eta_s, \xi_1, \ldots, \xi_s, \varsigma_1, \ldots, \varsigma_s, \tau_{11}, \ldots, \tau_{1s}, \ldots, \tau_{s1}, \ldots, \tau_{ss})$$
$$= \zeta + \sum_{j=1}^{s}\frac{\eta_j}{\xi_j}\varsigma_j + \sum_{j=1}^{s}\sum_{j'=1}^{s}\frac{\eta_j}{\xi_j}\frac{\eta_{j'}}{\xi_{j'}}\tau_{jj'}$$
$$- \sum_{j=1}^{s}\sum_{j'=1}^{s}(1-\phi_h(z_{j'}))\frac{\eta_j}{\xi_j}\tau_{jj'}$$

with $\zeta, \varsigma_j, \tau_{jj'}$ being $p$-dimensional and $\eta_j, \xi_j$ being real numbers. By the central limit theorem, the $\delta$-method and $A_{4n} = o_p(n_h^{-1/2})$ (proved in step 1), $\sqrt{n_h}[A_n - g(E(\phi(x_{hi}, \beta_0)))] = \sqrt{n_h}[g(\frac{1}{W_h}\sum_{i=1}^{n_h}w_{hi}\phi(x_{hi}, \beta_0)) - g(E(\phi(x_{hi}, \beta_0)))] + o_p(1) \to_d N(0, \Sigma_h)$, where $\Sigma_h$ is a $p \times p$ matrix. Since $A_n \to_p 0$, $g(E(\phi(x_{hi}, \beta_0))) = 0$ and the result follows.

**Step 3**. Show that $D_n = l_h''(\beta^\star)/W_h \to_p -U_h$, where $U_h$ is a positive definite matrix. Write $D_n = A_{5n} + A_{6n} + A_{7n}$, where

$$A_{5n} = \frac{1}{W_h} \sum_{i=1}^{r_h} w_{hi} (\log f_h)''(Y_{hi}, Z_{hi}, \beta^\star),$$

$$A_{6n} = \frac{1}{W_h} \sum_{i=1}^{r_h} w_{hi} \frac{\sum_{j=1}^{s} \frac{a_{hj}}{\hat{W}_h \hat{\pi}_{hj}} f_h''(Y_{hi}, z_j, \beta^\star)}{1 - \sum_{j=1}^{s} \frac{a_{hj}}{\hat{W}_h \hat{\pi}_{hj}} f_h(Y_{hi}, z_j, \beta^\star)},$$

$$A_{7n} = \frac{1}{W_h} \sum_{i=1}^{r_h} w_{hi} \frac{[\sum_{j=1}^{s} \frac{a_{hj}}{\hat{W}_h \hat{\pi}_{hj}} f_h'(Y_{hi}, z_j, \beta^\star)][\sum_{j=1}^{s} \frac{a_{hj}}{\hat{W}_h \hat{\pi}_{hj}} f_h'(Y_{hi}, z_j, \beta^\star)]^\tau}{[1 - \sum_{j=1}^{s} \frac{a_{hj}}{\hat{W}_h \hat{\pi}_{hj}} f_h(Y_{hi}, z_j, \beta^\star)]^2}.$$

Since

$$E_h \left\| \frac{1}{W_h} \sum_{i=1}^{r_h} w_{hi} (\log f_h)''(Y_{hi}, Z_{hi}, \beta^\star) - \frac{1}{W_h} \sum_{i=1}^{r_h} w_{hi} (\log f_h)''(Y_{hi}, Z_{hi}, \beta_0) \right\|$$

$$\leq \frac{n_h}{W_h} \max_{i \leq n_h} w_{hi} E_h \max_{\beta \in B_n} \|(\log f_h)''(y, z, \beta) - (\log f_h)''(y, z, \beta_0)\| \to 0,$$

$$A_{5n} \to_p \sum_{j=1}^{s} \phi_h(z_j) \pi_{hj}''(\beta_0)$$

$$- \sum_{j=1}^{s} \phi_h(z_j) \int \frac{f_h'(y, z_j, \beta_0)[f_h'(y, z_j, \beta_0)]^\tau}{f_h(y, z_j, \beta_0)} dF_h(y).$$

Under condition (iii), for $t = 1$ or $2$, since $\omega_0 \leq \sum_{j=1}^{s} \phi_h(z_j) f_h(Y_{hi}, z_j, \beta_0) \leq 1$ and $|\sum_{j=1}^{s} f_h(Y_{hi}, z_j, \beta^\star) c_{n,j}| \leq \sum_{j=1}^{s} |c_{n,j}|$, we have

(18)
$$\frac{1}{(1 - \sum_{j=1}^{s} \frac{a_{hj}}{\hat{W}_h \hat{\pi}_{hj}} f_h(Y_{hi}, z_j, \beta^\star))^t}$$
$$= \frac{1}{(\sum_{j=1}^{s} \phi_h(z_j) f_h(Y_{hi}, z_j, \beta_0))^t}(1 + o_p(1)),$$

uniformly in $i$. Then $A_{6n} \to_p \sum_{j=1}^{s} [1 - \phi_h(z_j)] \pi_{hj}''(\beta_0)$, and

$$A_{7n} \to_p \int \frac{\sum_{j=1}^{s} \phi_h(z_j) f_h'(y, z_j, \beta_0)[\sum_{j=1}^{s} \phi_h(z_j) f_h'(y, z_j, \beta_0)]^\tau}{\sum_{j=1}^{s} \phi_h(z_j) f_h(Y_{hi}, z_j, \beta_0)} dF_h(y).$$

Hence, $D_n \to_p -U_h$ with $U_h$ equals

$$\int \left[ \sum_{j=1}^{s} \frac{\phi_h(z_j) f_h'(y, z_j, \beta_0)[f_h'(y, z_j, \beta_0)]^\tau}{f_h(y, z_j, \beta_0)} \right.$$

$$\left. - \frac{\sum_{j=1}^{s} \phi_h(z_j) f_h'(y, z_j, \beta_0)[\sum_{j=1}^{s} \phi_h(z_j) f_h'(y, z_j, \beta_0)]^\tau}{\sum_{j=1}^{s} \phi_h(z_j) f_h(Y_{hi}, z_j, \beta_0)} \right] dF_h(y).$$

By Cauchy's inequality and condition (iv), $U_h$ is positive definite.

**Step 4**. Show that $P(l_h(\beta) - l_h(\beta_0) < 0$ for all $\beta \in \partial B_n) \to 1$ for each $h$. When $\beta \in \partial B_n$, $\beta = \beta_0 + n^{-1/3}u$ with $\|u\| = 1$. Then by (17) and results in step 3,

$$\frac{l_h(\beta) - l_h(\beta_0)}{W_h} = n^{-2/3}\left[n^{1/3}\frac{u^\tau l'_h(\beta_0)}{W_h} - \frac{1}{2}u^\tau U_h u + o_p(1)\right].$$

Let $\lambda$ be the smallest eigenvalue of $U_h$. Since $U_h$ is positive definite, $\lambda > 0$. Then

$$P\left(\left\|n^{1/3}\frac{u^\tau l'_h(\beta_0)}{W_h}\right\| \leq \frac{\lambda}{4}\right) \geq P\left(\left\|\frac{\sqrt{n_h}}{W_h}l'_h(\beta_0)\right\| \leq \frac{\lambda}{4}n^{-1/3}n_h^{1/2}\right) \to 1.$$

The result follows since $u^\tau U_h u/2 \geq \lambda/2$. We now prove the second conclusion in (8). It follows from step 2 of the previous proof that $\sqrt{n}l'(\beta_0) = \sum_{h=1}^H W_h \frac{\sqrt{n}}{\sqrt{n_h}} \frac{\sqrt{n_h}}{W_h} l'_h(\beta_0) \to_d N(0, \Sigma)$, where $\Sigma = \sum_{h=1}^H c_h^2 \Sigma_h/k_h$, $k_h = \lim_{n \to \infty} \frac{n_h}{n}$, and $c_h = \lim_{n \to \infty} N_h/N$ are defined in condition (i). Let $U = \sum_{h=1}^H c_h U_h$. It follows from step 3 of the previous proof that $l''(\beta_0) = \sum_{h=1}^H W_h \times l''_h(\beta_0)/W_h \to_p -U$. By Taylor's expansion and the fact that $l'(\hat{\beta}) = 0$, $-l'(\beta_0) = (\hat{\beta} - \beta_0)^\tau l''(\beta_0) + o_p(\|(\hat{\beta} - \beta_0)^\tau l''(\beta_0)\|)$. Then

(19) $$\hat{\beta} - \beta_0 = [-l''(\beta_0)]^{-1}l'(\beta_0) + o_p(\|\hat{\beta} - \beta_0\|).$$

Hence $\hat{\beta} - \beta_0 = [-l''(\beta_0)]^{-1}l'(\beta_0)\frac{1}{1+o_p(1)} = O_p(n^{-1/2})$. Then, by (19) and the result in step 2, $\sqrt{n}(\hat{\beta} - \beta_0) = [-l''(\beta_0)]^{-1}\sqrt{n}l'(\beta_0) + o_p(1) \to_d N(0, \Lambda)$, where $\Lambda = U^{-1}\Sigma U^{-1}$.

By (18), we can show that $\sum_{i=1}^{r_h} \hat{p}_{hi} \to_p 1$. Then $\sum_{h=1}^H W_h \sum_{i=1}^{r_h} \hat{p}_{hi} = 1 + o_p(1)$. Let $k_h(\beta) = \sum_{i=1}^{r_h} p_{hi}Y_{hi}$ and $t_h(\beta) = \sum_{i=1}^{r_h} p_{hi}$. Then

$$\hat{\bar{Y}} - \bar{Y} = \hat{\bar{Y}} - EY + o_p(n^{-1/2})$$

$$= \sum_{h=1}^H W_h[k_h(\hat{\beta}) - t_h(\hat{\beta})EY](1 + o_p(1)) + o_p(n^{-1/2})$$

$$= \sum_{h=1}^H W_h[k_h(\beta_0) - t_h(\beta_0)EY + (\hat{\beta} - \beta_0)^\tau(k'_h(\beta^\star) - t'_h(\beta^\star)EY)]$$

$$+ o_p(n^{-1/2}),$$

where $\beta^\star$ is between $\hat{\beta}$ and $\beta$. By (18), we can show that $k'_h(\beta^\star) \to_p c_{h1}$ and $t'_h(\beta^\star) \to_p c_{h2}$ for some constants $c_{hj}$. Let $c = \sum_{h=1}^H c_h(c_{h1} - c_{h2}EY)$. Then

$$\sqrt{n}(\hat{\bar{Y}} - \bar{Y}) = \sum_{h=1}^H W_h \frac{\sqrt{n}}{\sqrt{n_h}} \sqrt{n_h}\left[k_h(\beta_0) - t_h(\beta_0)EY + \frac{c^\tau U^{-1}l'_h(\beta_0)}{W_h}\right] + o_p(1).$$

Similarly to step 2 in the previous proof, we can write

$$k_h(\beta_0) - t_h(\beta_0)EY = h\left(\frac{1}{W_h}\sum_{i=1}^{n_h} w_{hi}\varphi(x_{hi},\beta_0)\right) + o_p(n_h^{-1/2})$$

for a function $h$ and a vector $\varphi$. Let $\bar{\phi}_h = \frac{1}{W_h}\sum_{i=1}^{n_h} w_{hi}\phi(x_{hi},\beta_0)$, $E_h\phi = E(\bar{\phi}_h)$, $\bar{\varphi}_h = \frac{1}{W_h}\sum_{i=1}^{n_h} w_{hi}\varphi(x_{hi},\beta_0)$ and $E_h\varphi = E(\bar{\varphi}_h)$. Recall that $l'_h(\beta_0)/W_h = g(\bar{\phi}_h) + o_p(n_h^{-1/2})$ and $g(E_h\phi) = 0$, where $\phi(x_{hi},\beta_0)$ and $g$ are defined in step 2 of the previous proof. By the central limit theorem and the $\delta$-method, $\sqrt{n_h}[h(\bar{\varphi}_h) + c^\tau U^{-1}g(\bar{\phi}_h) - h(E_h\varphi)] \to_d N(0,\sigma_h^2)$, where

$$
\begin{aligned}
\sigma_h^2 = {} & [h'(E_h\varphi), c^\tau U^{-1}g'(E_h\phi)] \\
& \times [d_h E_h(\varphi,\phi)(\varphi,\phi)^\tau - E_h(\varphi,\phi)E_h(\varphi,\phi)^\tau] \\
& \times [h'(E_h\varphi), c^\tau U^{-1}g'(E_h\phi)]^\tau
\end{aligned}
\tag{20}
$$

and $d_h$ is defined in condition (ii). Then

$$\sum_{h=1}^H W_h \frac{\sqrt{n}}{\sqrt{n_h}} \sqrt{n_h}[h(\bar{\varphi}_h) + c^\tau U^{-1}g(\bar{\phi}_h) - h(E_h\varphi)] \to_d N(0,\sigma^2), \tag{21}$$

where $\sigma^2 = \sum_{h=1}^H c_h^2 \sigma_h^2/k_h$. It follows from (21) that

$$\sum_{h=1}^H W_h\left[k_h(\beta_0) - t_h(\beta_0)EY + \frac{c^\tau U^{-1}l'_h(\beta_0)}{W_h}\right] \to_p \sum_{h=1}^H c_h h(E_h\varphi). \tag{22}$$

But the left-hand side of (22) equals $\sum_{h=1}^H W_h[E_hY - EY] + o_p(1) \to_p 0$. Therefore, $\sum_{h=1}^H c_h h(E_h\varphi) = 0$. Then, it follows from (21) and $W_h = c_h + o(n^{-1/2})$ that $\sqrt{n}(\hat{\bar{Y}} - \bar{Y}) \to_d N(0,\sigma^2)$. The proof for $\sqrt{n}(\hat{\bar{Y}}_j - \bar{Y}_j) \to_d N(0,\sigma_j^2)$ is similar. This shows (9) and completes the proof of Theorem 1.  □

LEMMA A.1. *Assume the conditions of Theorem 1. Let $x_{hi} = (\delta_{hi}, Y_{hi}, Z_{hi})$, $i = 1,\ldots,n_h$ and $\{x^*_{h1},\ldots,x^*_{hn_h}\}$ be a bootstrap sample. Assume that $\psi(x,\beta)$ is continuous in $\beta$ and $\|\psi(x,\beta)\|^2$ is bounded by an integrable function in a neighborhood of $\beta_0$; then $\frac{1}{W_h}\sum_{i=1}^{n_h} w^*_{hi}\psi(x^*_{hi},\hat{\beta})$ and $\frac{1}{W_h}\sum_{i=1}^{n_h} w_{hi}\psi(x_{hi},\hat{\beta})$ converge to $E_h\psi(x_{hi},\beta_0)$ in probability and*

$$\hat{\sigma}_\psi^{-1}\sqrt{n_h}\left[\frac{1}{W_h}\sum_{i=1}^{n_h} w^*_{hi}\psi(x^*_{hi},\hat{\beta}) - \frac{1}{W_h}\sum_{i=1}^{n_h} w_{hi}\psi(x_{hi},\hat{\beta})\right] \to_{d^*} N(0,I),$$

*where $\hat{\sigma}_\psi^2 = \text{Var}^*_h(\sqrt{n_h}\frac{1}{W_h}\sum_{i=1}^{n_h} w^*_{hi}\psi(x^*_{hi},\hat{\beta}))$ and $\text{Var}^*_h$ denotes the bootstrap variance conditional on the data in stratum $h$. Furthermore,*

$$\hat{\sigma}_\psi^2 \to_p d_h E_h[\psi(x_{hi},\beta_0)\psi(x_{hi},\beta_0)^\tau] - E_h[\psi(x_{hi},\beta_0)E_h\psi(x_{hi},\beta_0)^\tau],$$

which is the asymptotic variance of $\frac{\sqrt{n_h}}{W_h}\sum_{i=1}^{n_h} w_{hi}\psi(x_{hi},\beta_0)$, where $d_h$ is defined in condition (ii) of Theorem 1.

PROOF OF THEOREM 2. We follow the proof of Theorem 1. Let $B_n^* = \{\beta : \|\beta - \hat{\beta}\| \leq n^{-1/3}\}$ and $\partial B_n^* = \{\beta : \|\beta - \hat{\beta}\| = n^{-1/3}\}$. When $\beta \in \partial B_n^*$, there exists $\beta^*$ between $\beta$ and $\hat{\beta}$ such that $l^*(\beta) - l^*(\hat{\beta}) = (\beta - \hat{\beta})^\tau \sum_{h=1}^H l_h^{*\prime}(\hat{\beta}) + \frac{1}{2}(\beta - \hat{\beta})^\tau \sum_{h=1}^H l_h^{*\prime\prime}(\beta^*)(\beta - \hat{\beta})$.

**Step 1**. Show that $A_n^* = l_h^{*\prime}(\hat{\beta})/W_h \to_p 0$. Let $A_{jn}^*$ be the bootstrap analog of $A_{jn}$ in the proof of Theorem 1. Then $A_n^* = A_{1n}^* + A_{2n}^* + A_{3n}^* + A_{4n}^*$. By Lemma A.1, $A_{1n}^* \to_p P_h(\delta = 1)E_c f_h'(y,z,\beta_0)/f_h(y,z,\beta_0)$ and $A_{3n}^* \to_p \sum_{j=1}^s [(1-\phi_h(z_j))P_h(\delta=1)E_c \frac{f_h'(y,z_j,\beta_0)}{\sum_{j=1}^s \phi_h(z_j)f_h(y,z_j,\beta_0)}]$. Let $c_{n,j}^* = \frac{a_{hj}^*}{\hat{W}_h^* \hat{\pi}_{hj}^*} - (1-\phi_h(z_j))$. By Lemma A.1 and the $\delta$-method, $\sqrt{n_h}c_{n,j}^* = O_p(1)$. Hence, the same argument used in the proof of Theorem 1 leads to $A_{2n}^* = o_p(1)$ and $A_{4n}^* = o_p(n_h^{-1/2})$.

**Step 2**. Show that $\sqrt{n_h}(A_n^* - A_n) \to_{d^*} N(0, \Sigma_h)$. Using the notation in step 2 of the proof of Theorem 1, we have $A_n^* = g(\frac{1}{W_h}\sum_{i=1}^{n_h} w_{hi}^*\phi(x_{hi}^*, \hat{\beta})) + o_p(n_h^{-1/2})$. By Lemma A.1, the $\delta$-method and $g'(\frac{1}{W_h}\sum_{i=1}^{n_h} w_{hi}\phi(x_{hi}, \hat{\beta}))^\tau \hat{\sigma}_\phi^2 g'(\frac{1}{W_h}\sum_{i=1}^{n_h} w_{hi} \times \phi(x_{hi},\hat{\beta})) \to_p \Sigma_h$, we have

$$\sqrt{n_h}\left[g\left(\frac{1}{W_h}\sum_{i=1}^{n_h} w_{hi}^*\phi(x_{hi}^*, \hat{\beta})\right) - g\left(\frac{1}{W_h}\sum_{i=1}^{n_h} w_{hi}\phi(x_{hi}, \hat{\beta})\right)\right] \to_{d^*} N(0, \Sigma_h).$$

Then the result follows from $l_h'(\hat{\beta})/W_h = g(\frac{1}{W_h}\sum_{i=1}^{n_h} w_{hi}\phi(x_{hi},\hat{\beta})) + o_p(n_h^{-1/2})$.

**Step 3**. Show that $l_h^{*\prime\prime}(\beta^*)/W_h \to_p -U_h$, where $U_h$ is defined in the proof of Theorem 1. Note that $l_h^{*\prime\prime}(\beta^*)/W_h = A_{5n}^* + A_{6n}^* + A_{7n}^*$. When $\|f_h'(y,z_j,\beta)\|^3$ is bounded by an integrable function $G_j(y)$, $\max_{i \leq n_h}\|f_h'(Y_{hi}^*, z_j, \beta^{**})\| \leq \max_{i \leq n_h}\|f_h'(Y_{hi}, z_j, \beta^{**})\| \leq \max_{i \leq n_h} G_j^{1/3}(Y_{hi}) = o_p(n_h^{1/3})$, where $\beta^{**}$ is between $\beta^*$ and $\hat{\beta}$. For $t = 1$ or $2$, similarly to (18) we have

$$(23) \quad \frac{1}{(1 - \sum_{j=1}^s \frac{a_{hj}^*}{\hat{W}_h^* \hat{\pi}_{hj}^*} f_h(Y_{hi}^*, z_j, \beta^*))^t} = \frac{1}{(\sum_{j=1}^s \phi_h(z_j) f_h(Y_{hi}^*, z_j, \hat{\beta}))^t}(1 + o_p(1))$$

uniformly in $i$. The result follows from Lemma A.1 and similar arguments to step 3 in the proof of Theorem 1.

**Step 4**. Show that $P_*(l^*(\beta) - l(\hat{\beta}) < 0$ for all $\beta \in \partial B_n^*) \to_p 1$. The proof is similar to step 4 in the proof of Theorem 1, using the results established in steps 1–3.

This proves the first conclusion in (13). The proof of the second conclusion in (13) follows from Lemma 1 and the same argument in the proof of Theorem 1.

Let $k_h^*(\hat{\beta}^*)$ and $t_h^*(\hat{\beta}^*)$ be the bootstrap analogs of $k_h(\hat{\beta})$ and $t_h(\hat{\beta})$, respectively, defined in the proof of Theorem 1. By Lemma A.1 and (23), $\sum_{i=1}^{r_h^*} \hat{p}_{hi}^* = 1 + o_p(1)$. A similar argument to the proof of Theorem 1 yields

$$\sqrt{n}(\hat{\bar{Y}}^* - \hat{\bar{Y}}) = \sqrt{n}\sum_{h=1}^H W_h \left[ h\left(\frac{1}{W_h}\sum_{i=1}^{n_h} w_{hi}^* \varphi(x_{hi}^*, \hat{\beta})\right) \right.$$
$$+ c^\tau U^{-1} g\left(\frac{1}{W_h}\sum_{i=1}^{n_h} w_{hi}^* \phi(x_{hi}^*, \hat{\beta})\right)$$
(24)
$$\left. - h\left(\frac{1}{W_h}\sum_{i=1}^{n_h} w_{hi}\varphi(x_{hi}, \hat{\beta})\right)\right]$$
$$+ o_p(1).$$

Note that

$$\sqrt{n}\sum_{h=1}^H W_h c^\tau U^{-1} g\left(\frac{1}{W_h}\sum_{i=1}^{n_h} w_{hi}\phi(x_{hi}, \hat{\beta})\right)$$
$$= \sqrt{n}\sum_{h=1}^H W_h c^\tau U^{-1}\left[\frac{l_h'(\hat{\beta})}{W_h} + o_p(n_h^{-1/2})\right] = c^\tau U^{-1}\sqrt{n} l'(\hat{\beta}) + o_p(1) \to_p 0.$$

Applying Lemma A.1 and the $\delta$-method to (24), we can show that $\sqrt{n}(\hat{\bar{Y}}^* - \hat{\bar{Y}}) \to_{d^*} N(0, \sigma^2)$. Similar arguments can show that $\sqrt{n}(\hat{\bar{Y}}_j^* - \hat{\bar{Y}}_j) \to_{d^*} N(0, \sigma_j^2)$. □

PROOF OF THEOREM 3. The proofs for the mean imputation estimators are similar to that of Theorem 1. Conditional on the sample, the mean of the random imputation estimators is equal to the mean imputation estimators. Then the results for the random imputation estimators follow from those for the mean imputation estimators and Lemma 1 of Schenker and Welsh (1988). □

**Acknowledgments.** The authors thank one Associate Editor and two referees for helpful comments and suggestions.

F. Fang  
GE Consumer Finance  
1800 Cailun Road  
Zhangjiang High-Tech Park  
Shanghai 201203  
People's Republic of China  
E-mail: fang.fang2@ge.com

Q. Hong  
Eli Lilly and Company  
Lilly Corporate Center D/C 0734  
Indianapolis, Indiana 46285  
USA  
E-mail: hong_quan@lilly.com

J. Shao
Department of Statistics
University of Wisconsin
Madison, Wisconsin 53706
USA
E-mail: shao@stat.wisc.edu